\documentclass[%
   aps,% American Physical Society style
   prb,% Physical review B substyle
   reprint,% (p)reprint
   amsmath,%
   amssymb,%
   superscriptaddress,%
   showpacs,%
   showkeys,%
   floatfix,%
]{revtex4-1}

\usepackage{upgreek} % could conflict
\usepackage{graphicx}
\usepackage{color}
\usepackage{dcolumn} % for decimal point alignment

\newcommand{\refeaso}{\textit{RE}FeAsO}
\newcommand{\refeasoplus}{\textit{RE}FeAsO (\textit{RE}~=~La, Ce, Pr, Nd,
Sm, Gd, and Tb)}

\newcommand{\degree}{\ensuremath{^{\circ}}}

\newcommand{\wyfrac}[2]{\ensuremath{{#1}/{#2}}}

\newcommand{\etal}{\textit{et~al.}}
\newcommand{\xray}{X-ray}
\newcommand{\figref}[1]{Fig.~\ref{#1}}
\newcommand{\ffigref}[1]{\figref{#1}} % at line start
\newcommand{\tabref}[1]{Tab.~\ref{#1}}
\newcolumntype{d}[1]{D{.}{.}{#1}}

\begin{document}
\title{Structural trends from a consistent set of single-crystal data of
\textit{RE}FeAsO (\textit{RE}~=~La, Ce, Pr, Nd, Sm, Gd, and Tb)}

\author{F.~Nitsche}
\affiliation{Department of Chemistry and Food Chemistry, Technische
Universit\"{a}t Dresden, D-01062 Dresden, Germany}
\author{A.~Jesche}
\affiliation{Max Planck Institute for Chemical Physics of Solids, D-01187
Dresden, Germany} 
\author{E.~Hieckmann}
\affiliation{Institute of Applied Physics, Technische Universit\"{a}t
Dresden, D-01062 Dresden, Germany}
\author{Th.~Doert}
\affiliation{Department of Chemistry and Food Chemistry, Technische
Universit\"{a}t Dresden, D-01062 Dresden, Germany}
\author{M.~Ruck}
\affiliation{Department of Chemistry and Food Chemistry, Technische
Universit\"{a}t Dresden, D-01062 Dresden, Germany}
\affiliation{Max Planck Institute for Chemical Physics of Solids, D-01187
Dresden, Germany} 
\email{fabian.nitsche@mailbox.tu-dresden.de}

\date{\today}

\begin{abstract}
A new crystal growth technique for single-crystals of \textit{RE}FeAsO
(\textit{RE}~=~La, Ce, Pr, Nd, Sm, Gd, and Tb) using
NaI/KI as flux is presented. Crystals with a size up to
$300\,\upmu$m were isolated for single-crystal X-ray diffraction
measurements. Lattice parameters were determined by LeBail fits of X-ray
powder data against LaB$_6$ standard. A consistent set of structural data
is obtained and interpreted in a hard-sphere model. Effective radii for the
rare-earth metal atoms for \textit{RE}FeAsO are deduced. The relation of the intra-
and inter-plane distances of the arsenic atoms is identified as limiter of
the phase formation, and its influence on $T_\text{c}$ is discussed.  
\end{abstract}

\pacs{%
61.66.Fn, % Structure of specific crystalline solids - Inorganic compounds
74.62.Bf, % Transition temperature variations, phase diagrams - Effects of material synthesis, crystal structure, and chemical composition
74.70.Xa, % Superconducting materials other than cuprates - Pnictides and Chalkogenides
81.10.Dn  % Methods of crystal growth; physics and chemistry of crystal growth, crystal morphology, and orientation - Growth from solutions
}

\keywords{REFeAsO, single crystal, Fe-based superconductors, transport
properties, flux, crystal growth}

\maketitle
\section{Introduction}
The discovery of superconductivity in the system LaFePO(F) by Kamihara
\etal\cite{Kamihara2006} and the subsequent examination of various compounds
with a square iron-net substructure such as
\refeaso\ (\textit{RE}-1111 with \textit{RE}~$=$~rare-earth
metal),
$A$Fe$_2$As$_2$ ($A$-122 with $A =$~Ca, Sr, Ba and Eu),
$B$FeAs ($B$-111 with $B = $~Li, Na, K),
Fe\textit{Ch} (11 with \textit{Ch}~$=$~Se$_{1-x}$Te$_x$) 
revealed a large set of structures and options of doping to study
superconductivity. 

\begin{figure}
    \centering
    \includegraphics[bb=7 34 226 330,clip]{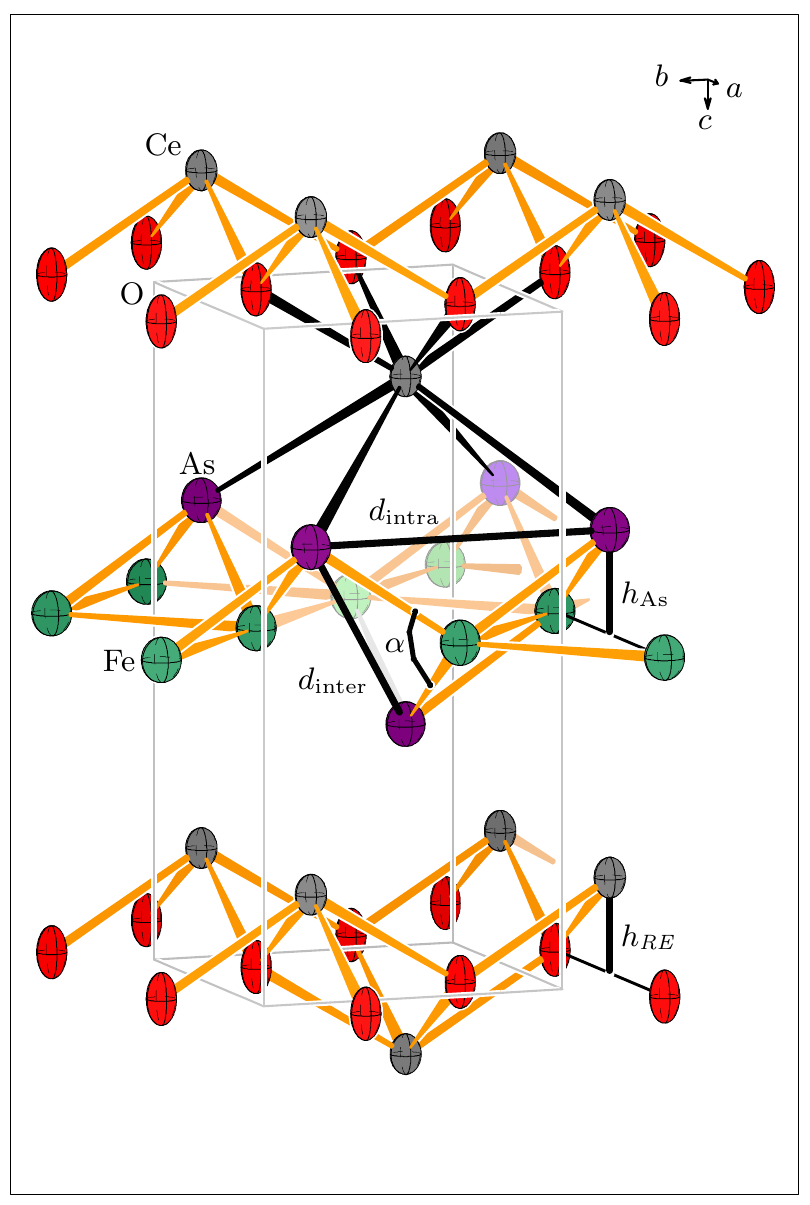}
\caption{\label{fig:refeaso} (color online) Crystal structure of CeFeAsO (space
group $P4/nmm$, no.~129). The displacement ellipsoids represent 95$\,$\%
localization probability. The black
bonds emphasize the square anti-prismatic
coordination of the rare-earth metal atom by four oxygen atoms and four
arsenic atoms. $h_\text{As}$ is the height of the
arsenic atoms above the plane of iron atoms.
$h_\text{\textit{RE}}$ is the height of the rare-earth metal atom above
the plane of oxygen atoms. $d_\text{intra}$ is the shortest intra-plane
distance between two arsenic atoms and equals the lattice parameter $a$.
$d_\text{inter}$ is the inter-plane distance between two arsenic atoms
above and below the iron atom net. The angle $\alpha$ is often referred to
as the tetrahedral angle and 
indicates the deviation from a tetrahedral coordination of the iron atom.} 
\end{figure}

In the \refeaso\ compounds, which crystallize in the ZrCuSiAs structure type
(\figref{fig:refeaso}), %
superconductivity can be achieved by electron\cite{Kamihara2008} and
hole\cite{Wen2008} doping as well as by applying
pressure\cite{Takahashi2009}. The underdoped compounds show a
tetragonal to
orthorhombic transition upon cooling followed by an anti-ferromagnetic
ordering.

\begin{figure}
    \centering
    \includegraphics{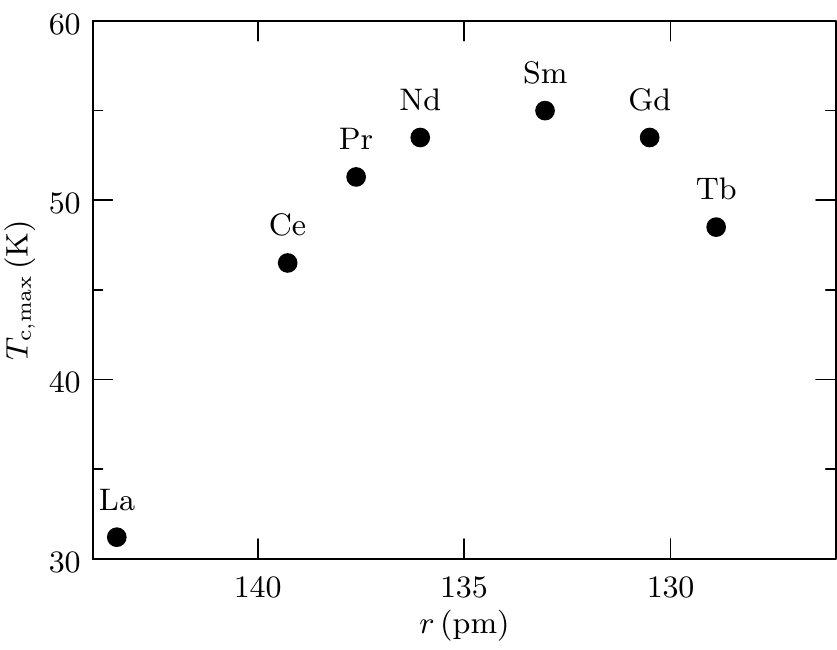}
\caption{\label{fig:tcmax}Evolution of the maximum $T_c$ for
\textit{RE}FeAsO$_{1-\delta}$\cite{Ren2008a,Yang2008,Yang2009} over the
effective rare-earth metal atom radii of \refeaso .}
\end{figure}

The impact of the rare-earth metal atom substitution on the maximal transition
temperature ($T_\text{c,max}$) achievable by doping is shown in
\figref{fig:tcmax}. 
The optimal starting point for electron doping thus seems to be SmFeAsO with
the highest achievable $T_\text{c}$ for \textit{RE}FeAsO$_{1-\delta}$. The
influence of the size of the rare-earth metal atom on the structure of
\refeaso\ and therby 
on superconductivity is essential for the understanding of the
mechanism of superconductivity as shown by Kuroki \etal\cite{Kuroki2009}, who
discussed the height of the arsenic atom above the iron atom layer
($h_\text{As}$) as a main influence parameter on superconductivity in
\refeaso.

Precise information on the evolution of the structures with respect to
rare-earth metal substitution and oxygen deficiency concentration is therefore
important. Structural information is often
obtained by Rietveld refinement of powder \xray\ diffraction data.
While this
method yields high quality cell parameters, structural refinement of
single-crystal \xray\ diffraction data offers a more accurate determination
of the structural parameters including displacement parameters.

Single-crystals of PrFeAsO and NdFeAsO of $70$--$100\,\upmu$m have been
grown from alkali metal chloride flux\cite{Quebe2000,Fang2009}. By
applying
pressure also SmFeAsO single-crystals of up to $150\,\upmu$m have been
obtained\cite{Zhigadlo2008,Karpinski2009}.  Using high
pressure and arsenic
as flux Ishikado \etal\ obtained large single-crystals
of PrFeAsO\cite{Ishikado2009}. From pellets of NdFeAsO and LaFeAsO synthesized
at high pressure Martin \etal\ have been able to isolate crystals with a
size up to half a millimeter\cite{Martin2009}. 
Ambient pressure crystal-growth 
yielding large single-crystals were done by Yan \etal\cite{Yan2009} for
LaFeAsO using NaAs and by Jesche \etal\cite{Jesche2009} for
CeFeAsO using Sn as flux. 
However, no structural investigation has
been conducted on a series of \textit{RE}-1111 obtained by the same
single-crystal growth technique.

Here we present a new ambient pressure method of single-crystal growth from
flux for \refeasoplus. Atomic
parameters were obtained by single-crystal \xray\ diffraction and
cell parameters were determined by LeBail fits of powder \xray\ diffraction
data against LaB$_6$ standard, yielding a consistent set of structural data.
Moreover, for the first time anisotropic displacement parameters for the
entire series of 
\textit{RE}-1111 are presented. In the light of these data, the influence of
the rare-earth metal atom substitution and electron doping on the structural
parameters is discussed.

\section{Experimental}
The starting materials were handled in an argon-filled glove box (M. Braun,
$p(\text{O}_2) \leq 1\,\text{ppm}$, $p(\text{H}_2\text{O}) \leq 1\,
\text{ppm}$, argon purification with molecular sieve and copper catalyst). 
Iron(II)-oxide powder ($99.9\,$\%, Sigma-Aldrich Chemie GmbH, iron content
checked by titration), arsenic
($99.999\,$\% Alfa Aeser GmbH \& Co.~KG) and the corresponding, freshly filed
rare-earth metal (lanthanum, cerium, praseodymium, neodymium: $99.9\,$\%,
Treibacher Industie AG; samarium: $99.9\,$\%, Chempur GmbH; gadolinium:
$99.9\,$\%,
ABCR GmbH \& Co.~KG; terbium: $99.9\,$\%, Acros Organics BVBA) were mixed and
transfered into glassy carbon crucibles. $300$--$500\,$wt.-\% of the eutectic
mixture of NaI/KI (both $99.5\,$\%, Gr\"ussing GmbH Analytika, dried at
$650\,$K in dynamic vacuum) was stacked on top of the reactants as flux.
Subsequently, the filled crucibles were sealed into silica ampoules under
dynamic vacuum. To prevent high arsenic vapor pressures, the ampules were
slowly heated to $1320\,$K within $24\,$h. An annealing period of three to six
days was applied, followed by slow cooling to $870\,$K with $1\,$K/h. After
quenching the ampules in air and removing the
flux with deionized water, plate-shaped
single-crystals of the \refeaso\ compounds suitable for structure and
transport investigations were obtained.

First attempts to grow single-crystals in accordance with previous studies on
\textit{RE}\textit{TM}$_x$As$_2$ (\textit{TM}~=~transition
metal)\cite{Rutzinger2009} were conducted with rare-earth metal oxide
(\textit{RE}$_2$O$_3$),
arsenic, and iron as starting materials and
alkali metal 
chlorides as flux.
As observed before\cite{Karpinski2009}, \textit{RE}OCl hindered the
formation of phase-pure samples and single-crystals of the target compounds.
The less stable oxide iodides (\textit{RE}OI) are not formed when using
sodium or potassium iodide.  The exchange of rare-earth metal oxide by
iron(III)-oxide
or better by
iron(II)-oxide as oxygen source, improved single-crystal growth. However,
the synthesis of \refeaso\ with rare-earth metals heavier than terbium
failed applying the NaI/KI flux method.

For single-crystal \xray\
diffraction smaller crystals were choosen, since larger crystals often
showed broad reflection profiles due to stacking faults
or mechanical stress as will be discussed later.  The single-crystals were
isolated and cleaned in inert oil to keep the mechanical stress to a
minimum. Subsequently, the crystals were sealed into glass
capillaries only immobilized by adhesion to the glass wall by residual oil.

Single-crystal \xray\ diffraction data was collected at 293(1)$\,$K on a
Bruker SMART diffractometer 
using a molybdenum \xray\ source and graphite monochromator (Mo~K$\alpha$).
Numerical absorption correction was applied using
SADABS\cite{Sheldrick2008}. Structure solution and
refinement was done with SHELXS and SHELXL\cite{Sheldrick2007}, respectively.
Further details of the crystal structure investigations may be obtained from
Fachinformationszentrum Karlsruhe, 76344 Eggenstein-Leopoldshafen,
Germany (fax: (+49)7247-808-666; e-mail: \verb+crysdata@fiz-karlsruhe.de+,
\verb+http://www.fiz-karlsruhe.de/+) on
quoting the CSD numbers 421998 (\textit{RE} = La), 421999 (Ce), 422000 (Pr), 422001
(Nd), 422002
(Sm), 422003 (Gd),  and 422004 (Tb).

$\omega$-scans of single reflections were done with
APEX2\cite{APEX2} by rotating 1\degree. The crystals were cooled with
a nitrogen gas stream using a Cryostream Controller 700 by
Oxford Cryosystems for low temperature $\omega$-scans. 

Powder \xray\ diffraction patterns were measured at 293(1)$\,$K on a Stadi~P
diffractometer (Stoe \& Cie., Darmstadt, Cu~$K\alpha_1$, Ge~monochromator). 
Lattice constants were refined by a LeBail pattern decomposition against
an internal LaB$_6$-standard using GSAS\cite{Larson2000} and
EXPGUI\cite{Toby2001}. Pseudo-Voigt profile functions with a model for axial
divergence were applied to fit the
measured data. 

Growth features, crystal composition, and flux incorporation were checked by
scanning electron microscopy and energy dispersive \xray\ spectroscopy
(EDX) at an Ultra55 FEG-microscope (Carl Zeiss NTS GmbH) with a XFlash
EDX-detector 4010 (SDD) assembling a Quantax 400 (Bruker AXS Microanalysis
GmbH). The EDX-detector was calibrated with germanium standard.

Electrical resistivity measurements were carried out in a standard four probe
geometry in the temperature range of
$1.8$--$300\,$K by using a commercial physical property measurement system
(PPMS, Quantum Design).
 
\section{Results}
Single-crystals suitable for \xray\ diffraction measurements were obtained
for all \refeasoplus. Larger crystals were observed especially for the
heavier rare-earth metals Gd and Tb. Whereas the size generally ranged from
$30$ to $300\,\upmu$m for all \textit{RE}-1111, a crystal of half a millimeter
size was obtained for TbFeAsO.

\begin{figure}
\includegraphics{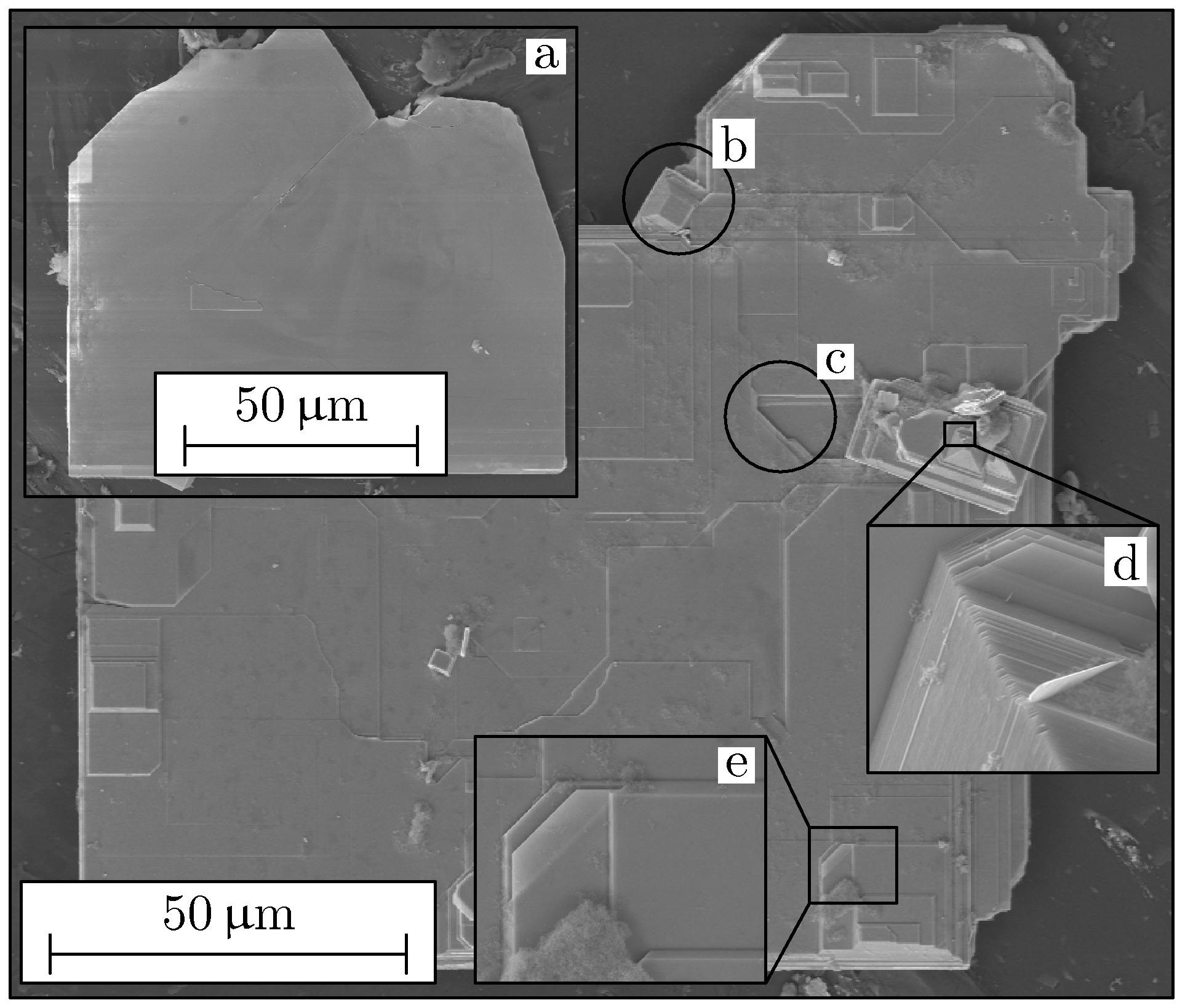}
\caption{\label{fig:sem}(a) Scanning electron microscope (SEM) image of 
a NdFeAsO crystal used for single-crystal \xray\ diffraction
studies. (b-e) SEM image of a GdFeAsO crystal showing common growth
instabilities of larger crystals:
(b)~additional crystal intergrown in random orientation, (c-e)~free grown
planes.}
\end{figure}

\ffigref{fig:sem} %
depicts the usual growth features of the single-crystals. The
crystal growth for \refeaso\ is generally much faster in the $ab$-plane
than along the $c$-axis.  Accordingly, the
platelets with a length of up to 300$\,\upmu$m are only few microns
thick rendering the crystals very fragile. Mechanical stresses can easily lead
to bending, splitting, or fracture of the crystal, which causes broad
reflection profiles in single-crystal \xray\ diffraction.
Additionally, the
larger crystals often show free growing planes (\figref{fig:sem}~c-e),
which can cause flux incorporation. Furthermore, intergrowth of crystals in
random orientation to the main crystal (\figref{fig:sem}~b) was observed
for larger crystals, hampering single-crystal \xray\ diffraction
investigations.  Smaller single-crystals (\figref{fig:sem}~a) are of better
quality, resulting in acceptable full width at half-maximum
of the Bragg reflections.

\begin{figure}
\includegraphics{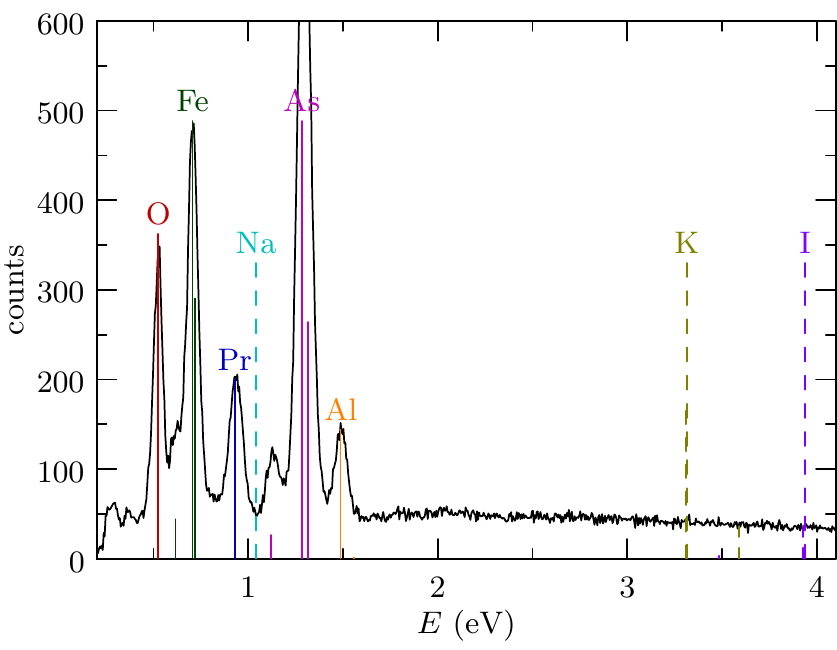}
\caption{\label{fig:edx}(color online) EDX spectrum of PrFeAsO.
Characteristic lines of the elements of the flux (positions marked by dashed
lines) are not observed. The
aluminum signal originates from the specimen holder.}
\end{figure}

Within the accuracy of the EDX measurement the
composition of the crystals was proved to be 1:1:1:1. Characteristic
emission lines of
neither potassium nor iodine were observed, thus substantial inclusion or
incorporation of the flux material can be excluded. Yet the
case of sodium is more complex, since the emission lines are often covered by
the signal of the rare-earth metals. However, in the case of PrFeAsO also
sodium could definitively be excluded, as can be seen in \figref{fig:edx}.

The parameters of single-crystal \xray\ diffraction data collections and the
results of the crystal structure refinements of the quarternary iron
pnictides are presented in \tabref{tab:refinedata}.%
\begin{table}
\caption{\label{tab:refinedata}Parameters of single-crystal \xray\
diffraction data collection at $293(1)\,$K and results of the structural
refinement of \refeaso.}
\begin{ruledtabular}
\begin{tabular}{%
   c% RE
   d{2.2}% 2theta max
   d{1.3}% R(int)
   d{1.3}% R(sigma)
   d{1.3}% R1
   d{1.3}% wR2
   d{1.2}% GooF
   c% Res. electron density
}
% Headings
   \multicolumn{7}{c}{}&$\Delta\rho_\text{max}$, $_\text{min}$\\
   \textit{RE}
   &\multicolumn{1}{c}{$\theta_\text{max}$ (\degree)}
   &\multicolumn{1}{c}{$R_\text{int}$}
   &\multicolumn{1}{c}{$R_\sigma$}
   &\multicolumn{1}{c}{$R_1$}
   &\multicolumn{1}{c}{$wR_2$}
   &\multicolumn{1}{c}{$S$}
   &(e$^-$\,\AA$^{-3}$)\\\hline
% Content
La & 37.52 & 0.037 & 0.016 & 0.027 & 0.023 & 2.22 & $1.69,\,-1.59$\\
Ce & 37.57 & 0.034 & 0.016 & 0.024 & 0.025 & 2.28 & $1.52,\,-1.87$\\
Pr & 37.65 & 0.039 & 0.017 & 0.025 & 0.018 & 1.66 & $2.01,\,-2.70$\\
Nd & 39.78 & 0.017 & 0.009 & 0.018 & 0.018 & 3.11 & $1.74,\,-1.15$\\
Sm & 37.57 & 0.024 & 0.010 & 0.011 & 0.013 & 1.92 & $0.89,\,-1.41$\\
Gd & 37.66 & 0.025 & 0.016 & 0.024 & 0.025 & 2.58 & $2.23,\,-2.25$\\
Tb & 37.52 & 0.018 & 0.009 & 0.012 & 0.016 & 2.51 & $1.20,\,-1.04$\\
\end{tabular}
\end{ruledtabular}
\end{table}
The lattice parameters from powder diffraction and atomic data
from single-crystal diffraction are compiled in
\tabref{tab:structdata}. %
\begin{table*}
\caption{\label{tab:structdata}Structural parameters of \refeaso\ in space group
 $P4/nmm$ (No.~129) and $Z=2$ measured at $293(1)\,$K. Cell parameters were
determined by LeBail fits of powder \xray\
diffraction patterns. Atomic positions and anisotropic displacement parameters
were refined from single-crystal \xray\ diffraction data.}
\begin{ruledtabular}
\begin{tabular}{%
   c% RE
   d{3.1}% Shannon radii
   d{3.5}% a
   d{3.5}% c
   d{1.8}% zRE
	d{5.0}% U11 = U22
	d{5.0}% U33
   d{1.8}% zAs
	d{5.0}% U11 = U22
	d{6.0}% U33
	d{5.0}% U11 = U22
	d{6.0}% U33
	d{6.0}% U11 = U22
	d{7.0}% U33
}
% Headings
   \multicolumn{4}{c}{}
   &\multicolumn{3}{l}{\textit{RE}%
       \footnote{%
          \textit{RE}, As: $2c$, $x = \wyfrac{1}{4}$, %
                                 $y = \wyfrac{1}{4}$, $z$; %
          Fe: $2b$, $x = \wyfrac{3}{4}$, $y = \wyfrac{1}{4}$, %
                    $z = \wyfrac{1}{2}$; %
           O: $2a$, $x = \wyfrac{3}{4}$, $y = \wyfrac{1}{4}$, $z = 0$}}
   &\multicolumn{3}{l}{As}
   &\multicolumn{2}{l}{Fe}
   &\multicolumn{2}{l}{O}\\
   \textit{RE} 
   & \multicolumn{1}{c}{$r$ (pm)\footnote{Radii of
	the corresponding trivalent rare-earth metal in octahedral coordination from
	Ref. \onlinecite{Shannon1976}}}
   & \multicolumn{1}{c}{$a$ (pm)} 
   & \multicolumn{1}{c}{$c$ (pm)} 
   & \multicolumn{1}{c}{$z$}
   & \multicolumn{1}{c}{$U_{11}$\footnote{All $U_{ij}$ in pm$^2$, $U_{11} =
	U_{22}$, $U_{12} = U_{23} = U_{13} = 0$}}
   & \multicolumn{1}{c}{$U_{33}$}
   & \multicolumn{1}{c}{$z$} 
   & \multicolumn{1}{c}{$U_{11}$}
   & \multicolumn{1}{c}{$U_{33}$}
   & \multicolumn{1}{c}{$U_{11}$}
   & \multicolumn{1}{c}{$U_{33}$}
   & \multicolumn{1}{c}{$U_{11}$}
   & \multicolumn{1}{c}{$U_{33}$}
	\\\hline
% Content
La & 130 & 403.67(1) & 872.18(4)
   & 0.14141(5) & 57(1) & 85(2) 
	& 0.65138(9) & 85(2) & 107(4) 
	& 88(3) & 106(5) 
	& 68(14) & 142(27)\\
Ce & 128.3 & 400.58(1) & 862.89(6)
   & 0.14106(4) & 50(1) & 85(2) 
	& 0.65442(8) & 79(2) & 102(3) 
	& 81(3) & 105(5) 
	& 46(12) & 145(24)\\
Pr & 126.6 & 398.89(1) & 859.66(8)
   & 0.13960(4) & 62(1) & 96(2) 
	& 0.65608(7) & 86(2) & 107(3) 
	& 87(2) & 121(4) 
	& 74(11) & 94(20)\\
Nd & 124.9 & 397.13(1) & 856.55(5)
   & 0.13899(3) & 61(1) & 93(1) 
	& 0.65725(6) & 85(2) & 107(3) 
	& 84(2) & 109(4) 
	& 69(10) & 120(18)\\ 
Sm & 121.9 & 394.69(2) & 849.65(6)
   & 0.13705(2) & 56(1) & 88(1) 
	& 0.66007(4) & 80(1) & 96(2) 
	& 78(2) & 108(3) 
	& 65(8) & 77(12)\\
Gd & 119.3 & 391.99(3) & 844.51(8)
   & 0.13570(5) & 69(1) & 98(2) 
	& 0.6623(1) & 92(2) & 109(5) 
	& 93(4) & 112(7) 
	& 24(14) & 97(32)\\
Tb & 118 & 390.43(3) & 840.8(1)
   & 0.13455(3) & 57(1) & 69(1) 
	& 0.66389(6) & 80(1) & 76(3) 
	& 76(2) & 100(4) 
	& 59(8) & 115(19)\\ 
\end{tabular}
\end{ruledtabular}
\end{table*}

\begin{figure}
\includegraphics{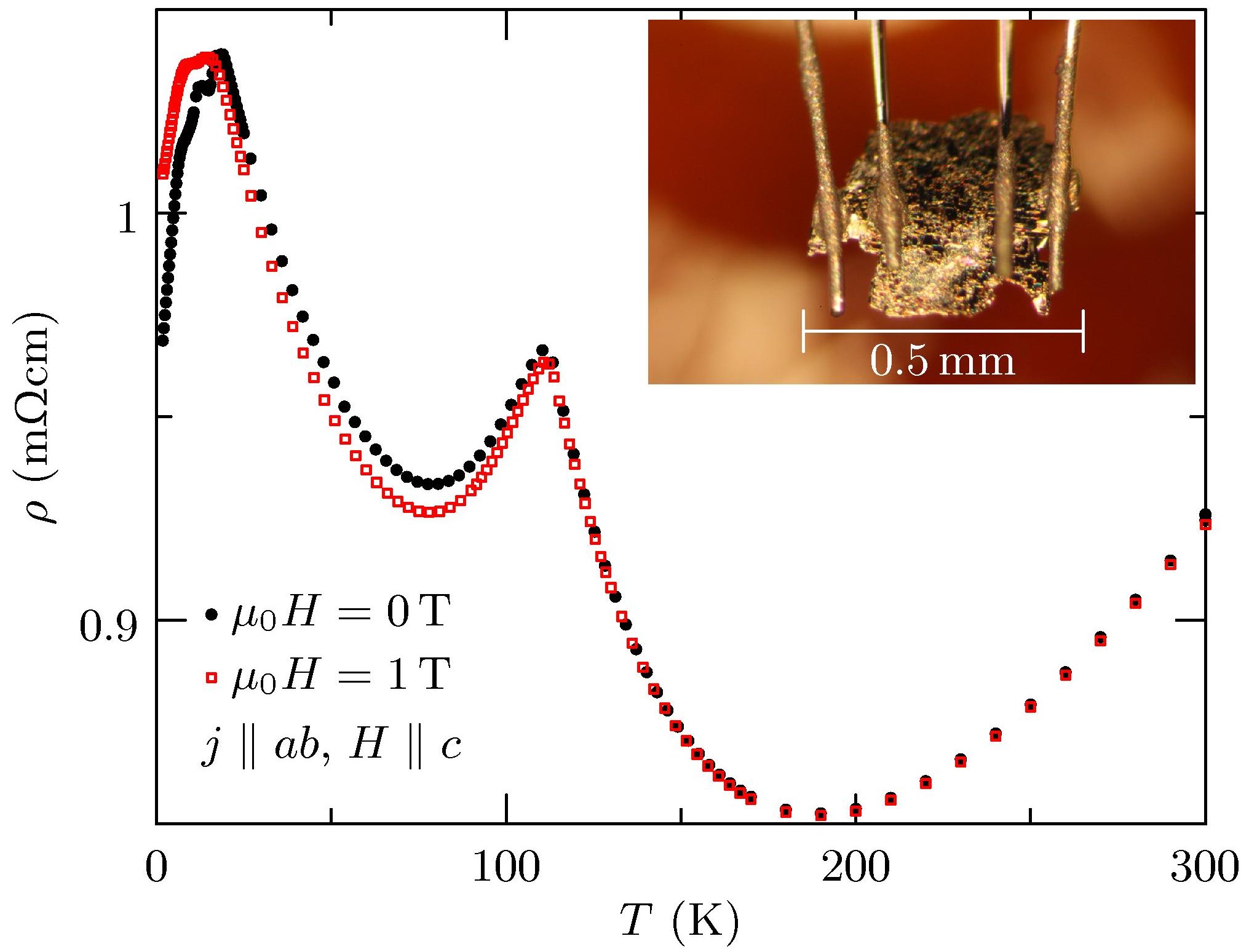}
\caption{\label{fig:tbfeaso}(color online) Temperature dependence of the electrical
resistivity of a TbFeAsO single-crystal (shown in the inset) in the
$ab$-plane. The local maximum at $T_\text{S} = 112\,$K results from a
structural transition.}
\end{figure}

Using NaI/KI-flux we obtained single-crystals, which are not only sufficient
for crystallographic investigations but also large enough to measure the
electrical resistivity along the basal plane with a standard four-probe
geometry (\figref{fig:tbfeaso}, inset). 
Over the
entire temperature range the resistivity of TbFeAsO varies only between
$0.85$
and $1.05\,$m$\Omega$cm. The room-temperature resistivity,
$\rho_{300\,\text{K}} \approx 1\,$m$\Omega$cm,
indicates poor metallicity. Upon cooling the resistivity of TbFeAsO passes a
minimum at about $180\,$K towards a local maximum centered at $T_\text{S} =
112\,$K, marking the structural transition and the magnetic ordering of iron,
which occures at a slightly lower temperature than reported for
polycrystalline material ($T = 122\,$K\cite{Luo2009}). 
Below $T = 75\,$K, $\rho(T)$
increases again, showing similarities to LaFeAsO\cite{Kamihara2008}, followed
by a decrease at $T < 18\,$K. The decrease of $\rho(T)$ changes slope
towards lower temperature with a linear temperature dependece at $T <
6\,$K. The onset of the anomaly at $T \approx 20\,$K is also observable in
polycrystalline material where no data is reported for $T \leq
19\,$K\cite{Luo2009}. A
similar behavior was published for NdFeAsO, where the change of the magnetic
structure of Fe at $T = 15\,$K results in a significant decrease in the
electrical resistivity of single-crystalline material with the current in the
$ab$-plane\cite{Tian2010}. The TbFeAsO single-crystal measured
(\figref{fig:tbfeaso}) has a
thickness of only $10\,\upmu$m and is accordingly sensitive to strain caused
by the wires and silver-paint contacts. Small deformation can
result in cracks along the sample, which might
be responsible for the observed differences to polycrystalline material.
Applying a magnetic field of $\mu_0 H = 1\,$T has no significant influence
on the structrual transition at $T_\text{S}$ and leads to a small negative
magneto-resistance at lower temperatures. The decrease in the electrical
resistivity occuring at $T = 18\,$K in zero field is shifted to $T = 14\,$K.

\begin{figure}
\includegraphics{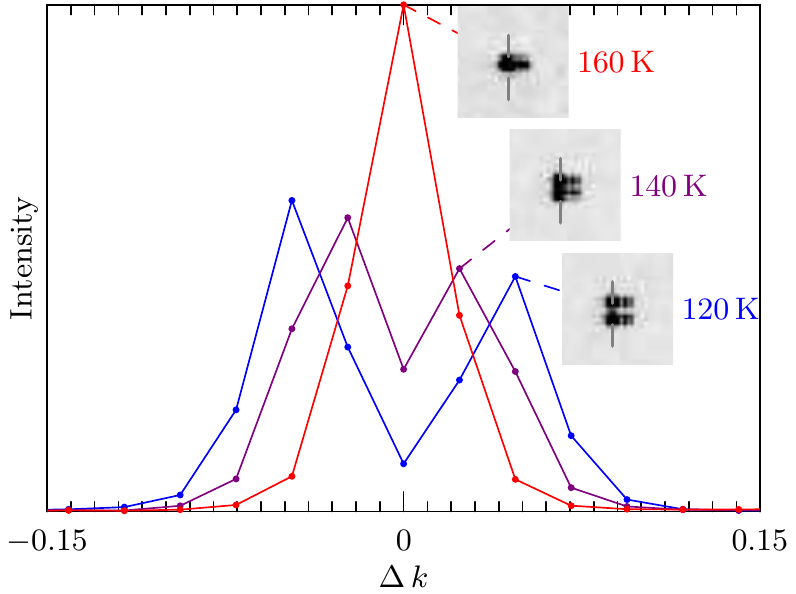}
\caption{\label{fig:splitting}(color online) $\omega$-scan of the Mo
K$\alpha_1$ and K$\alpha_2$ Bragg reflections $800$ of
LaFeAsO showing a splitting on cooling below the tetragonal to orthorhombic
transition. Lines are guides for the eye. The data points correspond to the
resolution of the area detector.}
\end{figure}

Splitting of the tetragonal Bragg reflection $800$ occuring below $160\,$K
 is demonstrated exemplarily for LaFeAsO (\figref{fig:splitting}), 
indicating the structural transition from space group $P4/nmm$ to $Cmme$ as
expected for undoped LaFeAsO.\cite{delaCruz2008}

\section{Discussion}
To assess the influence of the rare-earth metal substitution on the
structure of \textit{RE}-1111 and thereby on superconductivity, a suitable
scale has to be defined. Often, the radii of the substituted atoms are used
since a linear correlation to the volume of the structure is expected in
analogy to Vegard's law. 

As depicted in \figref{fig:refeaso}, the rare-earth metal atom is
coordinated by four oxygen atoms and four arsenic atoms.  This suggests the
use of the radii for eight-fold coordinated
\textit{RE}$^{3+}$ ions as revised by Shannon\cite{Shannon1976} and
originally deduced by Greis and Petzel\cite{Greis1974} from the nine-fold
coordination in \textit{RE}F$_3$ where the coordination polyhedron is
assembled by equal anions at almost equal distances.
 
\begin{figure}
\centering
\includegraphics{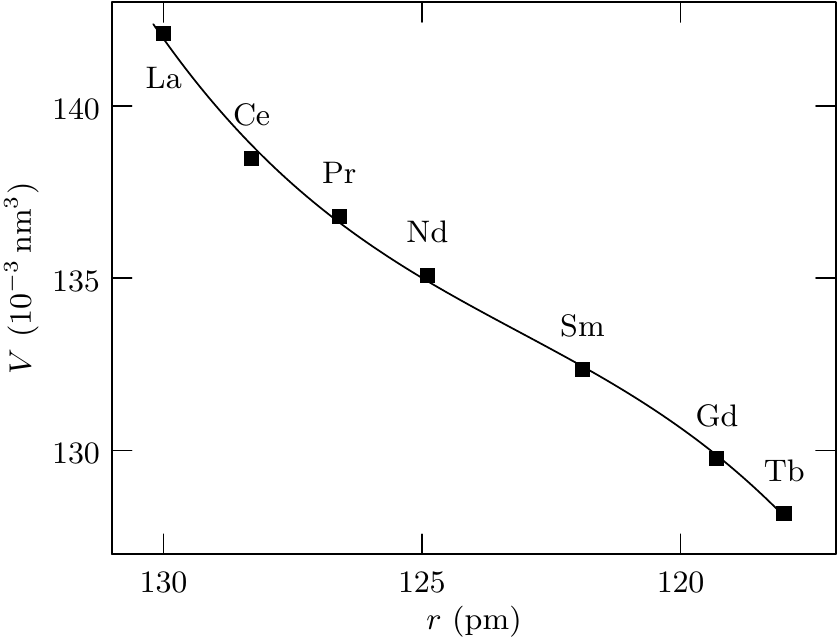}
\caption{\label{fig:volume}Unit cell volume of \refeaso\ over the rare-earth
metal radii as compiled by Shannon\cite{Shannon1976} for eight-fold
coordinated trivalent ions. The errorbars are within the size of the
symbols.}
\end{figure}

However, in \refeaso\ the square spanned by the four arsenic atoms is twice as
large as the square spaned by the oxygen atoms, leading to a strong distortion
of the square anti-prism with a large difference in cation-anion distances
($\bar{d}_\text{\textit{RE}-O}\approx 230\,$pm,
$\bar{d}_\text{\textit{RE}-As}\approx 330\,$pm). The more inhomogeneous 
coordination in \refeaso\ compared to the coordination in \textit{RE}F$_3$
creates a deviation from linearity of the volume evolution over the choosen
rare-earth metal radii in eight-fold coordination as shown 
in
\figref{fig:volume}.

\begin{figure}
\centering
\includegraphics{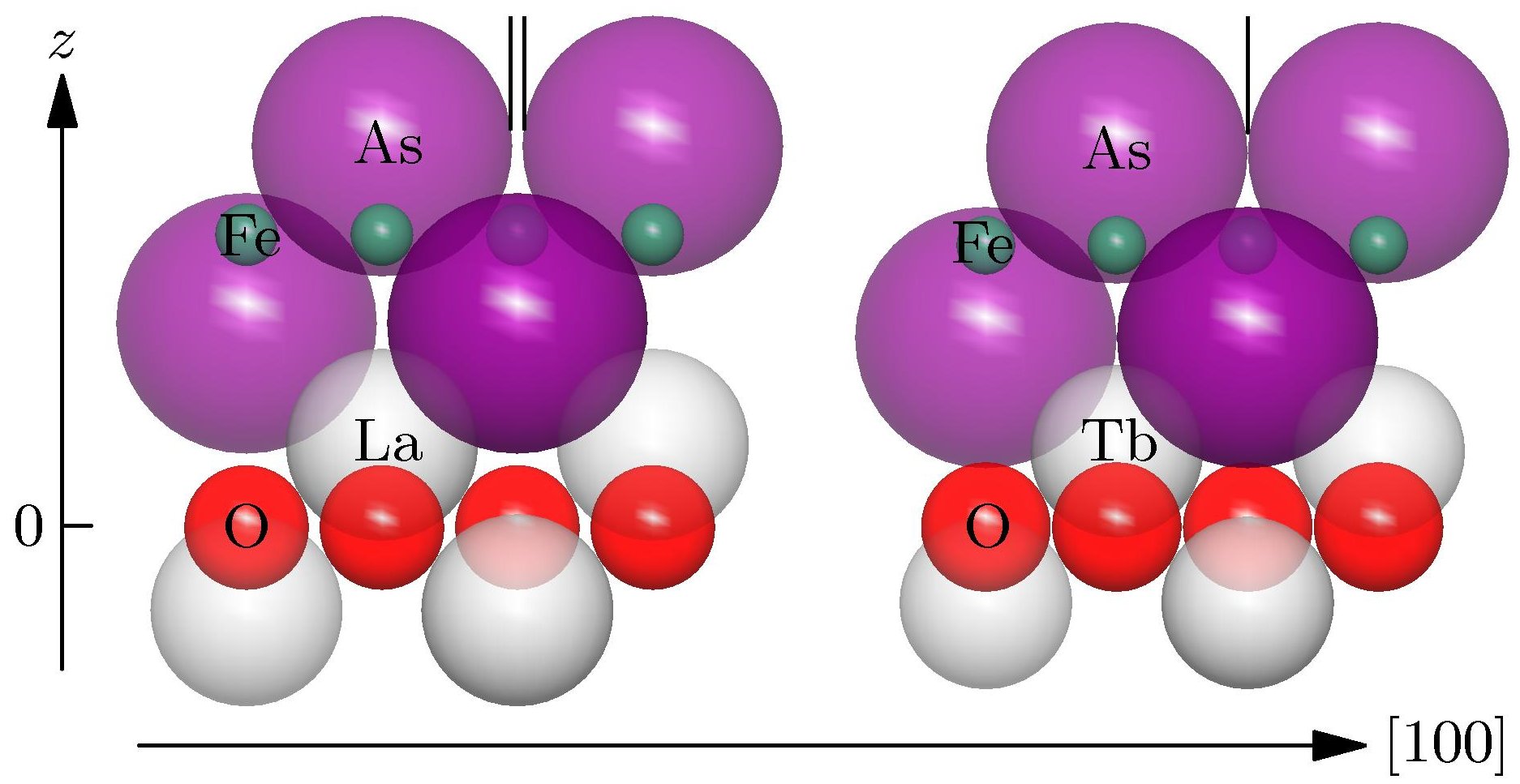}
\caption{\label{fig:hsexample}(color online) Hard-sphere model representation of LaFeAsO
(left) and TbFeAsO (right). The intra-plane distance (see
\figref{fig:refeaso}) of the arsenic atoms
approaches the inter-plane distance going from lanthanum to terbium.}
\end{figure}

Thorough investigation of the crystal structures of
\textit{RE}-1111 reveals, that the inter-plane distance of the arsenic
atoms (\figref{fig:refeaso}) is constant(!), $\bar{d}_\text{inter} =
389.4(4)\,$pm. It appears to be the minimum distance arsenic atoms can
attain in undoped \refeasoplus. The intra-plane distance of the arsenic
atoms, $d_\text{intra}$ (\figref{fig:refeaso}), is equal to the
$a$-axis, and longer than $d_\text{inter}$. Going from lanthanum
to terbium as depicted in \figref{fig:hsexample},
the intra-plane distance 
approaches the inter-plane distance of the arsenic atoms preventing a further
shrinkage of the structure. Consequently, all \refeaso\ compounds with
smaller rare-earth metal atoms (\textit{RE} = Dy, Ho, ...) have to be
synthesized under external pressure as
described in literature \cite{Quebe2000,Bos2008,Rodgers2009}. 

The iron-arsenic layer in TbFeAsO can thus be considered as a section of a
body centered cubic packing of arsenic atoms with iron atoms situated in the
tetrahedral voids. 

With $d_\text{inter}$ beeing constant over the rare-earth substitution
and the interaction between \textit{RE}, As and O beeing predominantly
ionic, as argued by Blanchard \etal\cite{Blanchard2010},
the structure of \refeaso\ can be analyzed in a hard-sphere model
(\figref{fig:hsexample}). The radius 
of the arsenic sphere is $r_\text{As} = \frac{1}{2}\,d_\text{inter}$.
The radius of the
iron sphere is deduced from the shortest inter-atomic distance of iron and
arsenic atoms
$r_\text{Fe} = d_\text{As-Fe}-r_\text{As}$. 
The
same can be done for the radius of the rare-earth metal sphere
($r_\text{\textit{RE}} = d_\text{As-\textit{RE}}-r_\text{As}$).  The radius
of the oxygen sphere cannot be determined directly from $d_\text{inter}$
and has to
be calculated from the radius of the rare-earth metal sphere and the
shortest inter-atomic distance of oxygen and rare-earth metal atom 
($r_\text{O} = d_\text{\textit{RE}-O}-r_\text{\textit{RE}}$).

The resulting effective radii are listed in
\tabref{tab:hardsphere}. %
\begin{table}
\caption{\label{tab:hardsphere}Calculated effective radii of the atoms in
\refeaso\ as deduced from the hard-sphere model. The errors are estimated by
combined variances.}
\begin{ruledtabular}
\begin{tabular}{%
   c% RE
   d{3.5}% rRE
   d{2.5}% rFe
   d{3.5}% rAs
   d{3.5}% a/2
   d{2.5}% rO
}
% Headings
   \textit{RE}
   &\multicolumn{1}{c}{$r_\text{\textit{RE}}$ (pm)}
   &\multicolumn{1}{c}{$r_\text{Fe}$ (pm)}
   &\multicolumn{1}{c}{$r_\text{As}$\footnote{$r_\text{As}=\frac{1}{2}\,d_\text{inter}$} (pm)}
   &\multicolumn{1}{c}{$\frac{1}{2}\,a$\footnote{$\frac{1}{2}\,a =
\frac{1}{2}\,d_\text{intra}$} (pm)}
   &\multicolumn{1}{c}{$r_\text{O}$ (pm)}\\
% Content
La & 143.42(9) & 46.76(7) & 194.42(6) & 201.84(1) & 93.12(9) \\
Ce & 139.28(8) & 46.11(7) & 194.46(5) & 200.29(1) & 95.10(8) \\
Pr & 137.62(8) & 45.72(6) & 194.66(5) & 199.45(1) & 95.15(8) \\
Nd & 136.06(6) & 45.37(5) & 194.57(4) & 198.57(1) & 95.45(6) \\
Sm & 133.17(5) & 44.81(4) & 194.86(3) & 197.35(1) & 95.97(5) \\
Gd & 130.5(1)  & 44.2(1)  & 194.94(7) & 196.00(2) & 96.5(1) \\
Tb & 128.90(7) & 43.90(6) & 195.05(5) & 195.22(2) & 96.73(8) \\
\end{tabular}
\end{ruledtabular}
\end{table}
According to the hard-sphere model, all radii except those of the rare-earth
metal atoms should be almost constant. Indeed, the deviations in the
calculated values are small, yet exhibit the limits of the model,
which does not
account for
polarization effects or covalent bonding. Nevertheless, the changes of
the rare-earth metal atom radii ($r_\text{\textit{RE}}$) are an order of
magnitude larger then those of the other atoms. The above mentioned
geometrical limit of the \refeaso\ series with $\text{\textit{RE}} =
\text{Tb}$ can
clearly be seen by the equalization of 
$d_\text{intra} \equiv a$ and 
$d_\text{inter} \equiv 2\,r_\text{As}$

\begin{figure}
\centering
\includegraphics{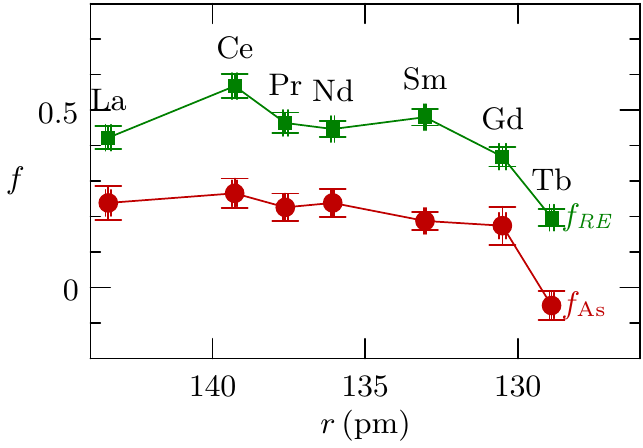}
\caption{\label{fig:anisofvib}(color online) Anisotropy of vibration $f$ of the arsenic and
rare-earth atoms in \refeaso\ with $f=0$ indicating an isotropic vibraional
behavior. Connecting lines are guides for the eye. Errorbars by combined
variances.} 
\end{figure}

Additionally, the sphere packing can be checked by the anisotropy of vibration 
\[
f=\frac{U_{33}-U_{11}}{\left(2\,U_{11}+U_{33}\right)/3}\text{,}
\]
which indicates the elongation of the displacement elipsoid along the
$c$-axis, with $f=0$ referring to an isotropic vibration. For all \textit{RE}
the arsenic atom shows almost the same anisotropy of vibration
(\ffigref{fig:anisofvib}). 
Only for the densely packed situation of TbFeAsO isotropic displacement is
observed. For the vibration of the heavier rare-earth atoms the trend
towards diminished anisotropy is observed, too.

\begin{figure}
\centering
\includegraphics{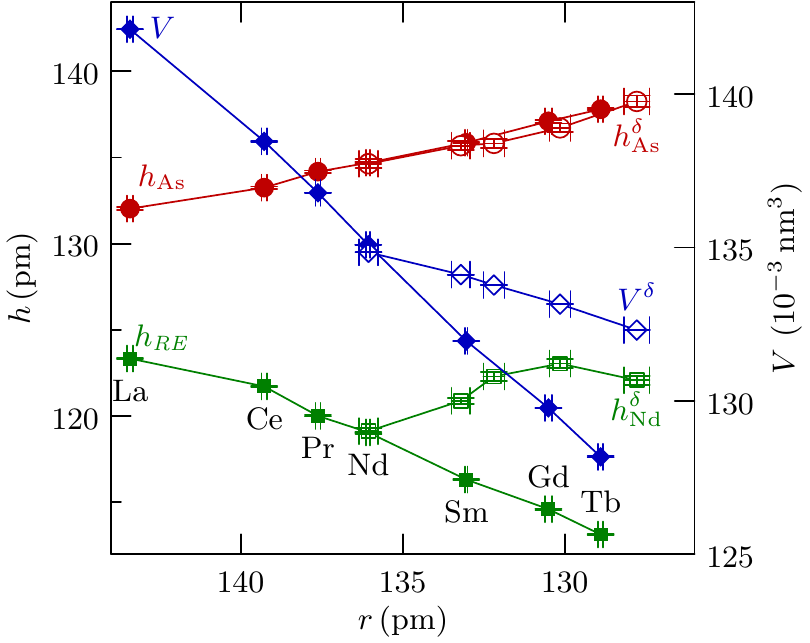}
\caption{\label{fig:height_volume}(color online)The height of the arsenic
atom ($h_\text{As}$ see \figref{fig:refeaso}) over the iron plane, the
height of the rare-earth metal atom ($h_\text{\textit{RE}}$ see
\figref{fig:refeaso}) over the oxygen layer and the volume $V$ over the
effective radii of the trivalent rare-earth metal deduced form the hard-sphere
model. The open symbols represent data for
NdFeAsO$_{1-\delta}$\cite{Lee2008}. From left to right: 
$\delta=0.05(1),$~RT, non-superconducting; 
$\delta=0.080(9),$~RT, $T_\text{c} = 35\,$K; 
$\delta=0.14(1),$~RT, $T_\text{c} = 44\,$K; 
$\delta=0.17(1),$~RT, $T_\text{c} = 51\,$K; 
$\delta=0.17(1),$~$10\,$K, $T_\text{c} = 51\,$K. Connecting lines are guides
for the eye. Errorbars by combined variances.}
\end{figure}

Using the effective radii of \textit{RE} as a scale to compare the
influence of the rare-earth metal substitution on the structure of
\textit{RE}-1111 leads to
a linear
evolution of structural features such as the
volume or the height of the arsenic atom over the square iron-net 
(\figref{fig:height_volume}). %
Although LaFeAsO fits the linear evolution of the volume of \refeaso,
it deviates slightly from the linear trends in the height of the arsenic atom
over the square iron-net 
($h_\text{As} = \left[z_\text{As}-0.5\right]\,c$) 
and in the 
height of the rare-earth metal atom 
over 
the oxygen layer 
($h_\text{RE} = z_\text{RE}\,c$). 
Since the radius for the rare-earth metal atom is deduced from the
inter-atomic distance of arsenic and the rare-earth metal, closer
coordination towards oxygen than expected from linear extrapolation results
in a comparetively large effective radius for the $f^0$-cation La$^{3+}$
(\tabref{tab:hardsphere}).  This is consistent with the larger decrease of
$T_\text{c,max}$ from
CeFeAsO$_{1-\delta}$ ($46.5\,$K) 
to LaFeAsO$_{1-\delta}$
($31.2\,$K) 
compared to the decrease from 
PrFeAsO$_{1-\delta}$ ($51.3\,$K)\cite{Ren2008a} to 
CeFeAsO$_{1-\delta}$
as can be seen in \figref{fig:tcmax}.

\begin{figure}
\centering
\includegraphics{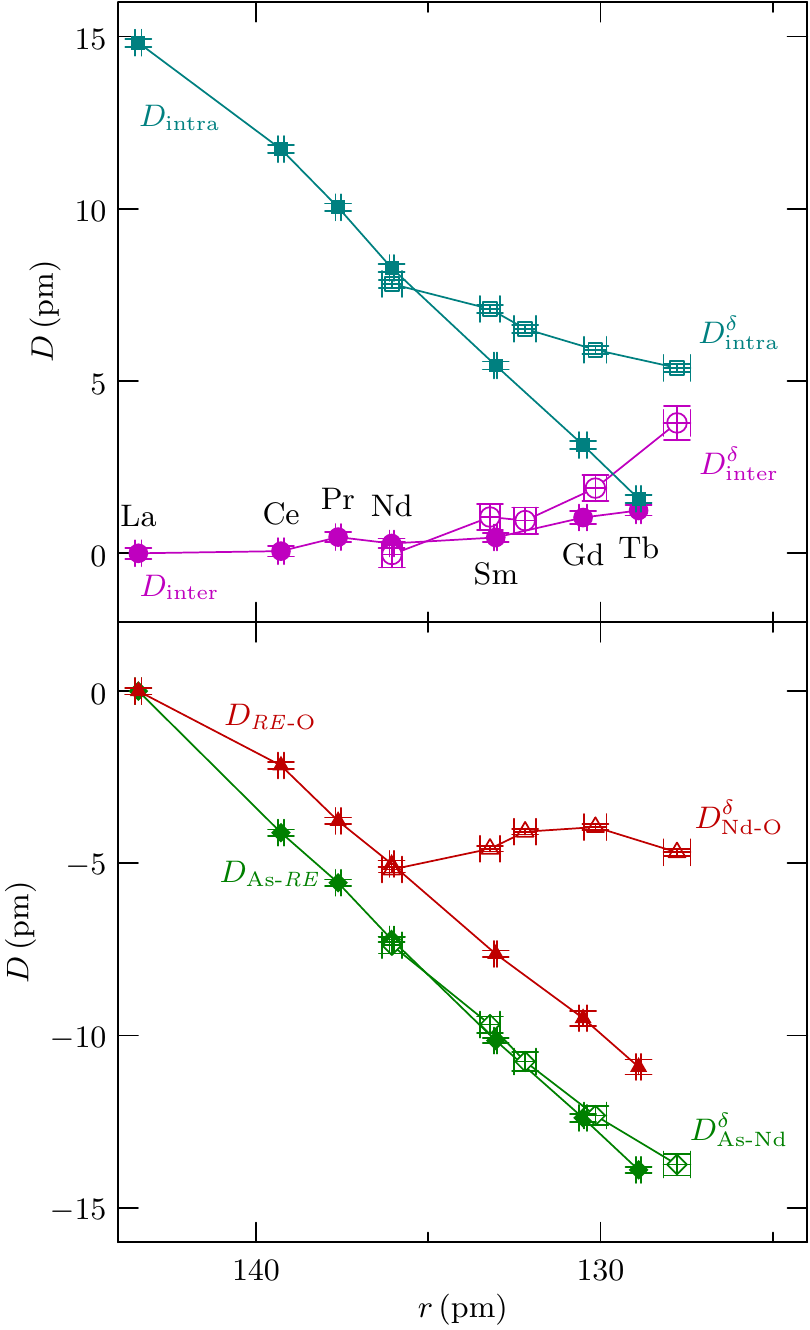}
\caption{\label{fig:distances}(color online) Plot of the normalized inter-atomic distances 
$D_\text{A-B} = 
d^\text{\textit{RE}FeAsO}_\text{A-B} -
 d^\text{LaFeAsO}_\text{A-B}$ 
with respect to LaFeAsO over the
calculated effective rare-earth atom radii deduced form the hard-sphere
model. For better comparison of the normalized intra- and inter-plane
distance of the arsenic atoms, $D_\text{intra}$ is calculated by 
$d^\text{\textit{RE}FeAsO}_\text{intra} -
 d^\text{LaFeAsO}_\text{inter}$.
The open symbols represent data for
NdFeAsO$_{1-\delta}$\cite{Lee2008}. From left to right: 
$\delta=0.05(1),$~RT, non-superconducting; 
$\delta=0.080(9),$~RT, $T_\text{c} = 35\,$K; 
$\delta=0.14(1),$~RT, $T_\text{c} = 44\,$K; 
$\delta=0.17(1),$~RT, $T_\text{c} = 51\,$K; 
$\delta=0.17(1),$~$10\,$K, $T_\text{c} = 51\,$K. Connecting lines are guides
to the eye. Errorbars by combined variances.}
\end{figure}

The exceptional position of LaFeAsO can also be observed in the normalized
inter-atomic distances $D$ as depicted in \figref{fig:distances}.%
The largest changes in inter-atomic distances are observed for the
interactions directly influenced by the rare-earth metal substitution: 
$d_\text{\textit{RE}-As}$, $d_\text{\textit{RE}-O}$ and
$d_\text{intra}$ decrease monotonically, with the latter
corresponding to the $a$-parameter. The inter-atomic distances between iron
and arsenic (not shown in the Figure) and the inter-plane
distance of the arsenic atoms are almost
invariant to rare-earth metal substitution.

To assess the influence of oxygen vacancy concentration on the structure of
\refeaso, literature data for NdFeAsO$_{1-\delta}$ from Lee
\etal\cite{Lee2008}
were also evaluated with the hard-sphere model (open symbols in
\figref{fig:height_volume} and \figref{fig:distances}).
Oxygen deficiency weakens the bonding in the NdO layer. The
distance between oxygen and neodymium atoms as well as $h^\delta_\text{Nd}$
increases slightly. The weaker Nd-O bonding, respectively the higher
charge of the 
(NdO$_{1-\delta}$)$^{(1+2\,\delta)+}$ and 
(FeAs)$^{(1+2\,\delta)-}$ layers is compensated by stronger Nd-As bonding.
The shorter $d^\delta_\text{Nd-As}$ results in smaller $r_\text{Nd}$.
Concurrently, the distance of arsenic atoms to the iron layer
$h^\delta_\text{As}$ is
enlarged. As the inter-plane As-As distance increases, the geometrical
limit of the structure type is changed. Further contraction (here due to
cooling
from RT to $10\,$K) leads again to the equalization of 
$d_\text{inter}$ and 
$d_\text{intra}$, however at a value that is about $4\,$pm larger.

As depicted by Lee
\etal\cite{Lee2008},  
the angle $\alpha$ of the doped \refeaso\
compounds at low temperatures is
a reasonable indicator for $T_\text{c,max}$, 
which is highest for the tetrahedral angle
of $109.5\degree$. However, the
structural arguments do not suffice. Taking LaFeAsO$_{1-x}$F$_x$ as an
example, doping
triggers superconductivity at $x\approx 0.05$, with $T_\text{c,max} \approx
24\,$K at $x\approx 0.1$, and a subsequent suppression of superconductivity
at higher doping levels\cite{Luetkens2009}. Geometrically, the
increase in doping should further increase $\alpha$ towards the ideal
tetrahedral coordination for the iron atoms, with a further increase in
$T_\text{c}$. It can thus be argued that high-$T_\text{c}$ values originate from
a coincidence of electronical and structural optima.

Comparing the influence of rare-earth substitution exemplarily with the data of
oxygen deficiency doping data of Lee \etal\ shows the high impact of
electronical doping on the structure. Kuroki \etal\cite{Kuroki2009}
indicated the strong influence of $h_\text{As}$ on the electronical
structure of \textit{RE}-1111. It was predicted that the combination of
large cell volume $V$ and increased height of the arsenic atom $h_\text{As}$
results in high $T_\text{c}$. As can be deduced from the diametric trends in
\figref{fig:height_volume}, a medium sized rare-earth metal atom in
combination with a large oxygen deficiency should be used. In accordance
with experimental data, NdFeAsO$_{1-\delta}$ seems to be close to the optimum.

\section{Conclusions}
Using alkali metal iodides as flux facilitated the free growth of \refeasoplus\
single-crystals suitable for resistivity measurements and single-crystal \xray\
diffraction studies. The elaborated consistent set of structural data for the
\textit{RE}-1111 compounds and the comparison with structural data of
NdFeAsO$_{1-\delta}$ from literature 
allowed a thorough
investigation of the influence of the rare-earth metal substitution and
doping on the structural features of \textit{RE}-1111. 

A hard-sphere model proved to be suitable for the interpretation of structural
trends. The thereby revealed geometrical limit for the structure type
rationalizes the failure of
ambient pressure synthesis of heavier \textit{RE}-1111 then TbFeAsO.

Additionally, an interplay of electronical and structural
optima for high-$T_\text{c}$ values is proposed.  

The structural data also provide a reliable basis for theoretical
investigations of the electronic structure of the undoped compounds and
their relation to their superconducting congeners. 

Further single-crystal studies of
doped \refeaso\ compounds and the evolution over the temperature will help
to complete the geometrical model of the iron-based superconductors.

\section{Acknowledgment}
The authors like to thank Jutta Krug, Alexander Gerisch, and Tilmann Meusel
for their help with sample preparation and measurements.

\end{document}